\documentclass[conference,10pt]{IEEEtran}
\usepackage{graphicx}
\usepackage{amsmath}
\usepackage{mathtools}
\usepackage{amssymb}
\usepackage{math_symbols}
\usepackage{calc}
\usepackage{booktabs,hhline} 
\usepackage{enumitem}
\usepackage{siunitx}
\usepackage{acronym}
\usepackage[english]{babel}
\usepackage{times}
\usepackage{url}
\usepackage{tikz}
\usetikzlibrary{calc,arrows,positioning}
\usepackage{tabularx}
\acrodef{ssm}[SSM]{state-space model}
\acrodef{ss}[SS]{state-space}
\acrodef{brr}[BRR]{Bayesian recursive relation} 
\acrodef{pdf}[PDF]{probability density function}
\acrodef{cke}[CKE]{Chapman-Kolmogorov equation}
\acrodef{mc}[MC]{Monte Carlo}
\acrodef{rmse}[RMSE]{root mean-squared error}
\acrodef{mse}[MSE]{mean-squared error}
\acrodef{nis}[NIS]{normalized innovation squared}
\acrodef{anees}[ANEES]{average normalized estimate error squared}
\acrodef{kld}[KLD]{Kullback-Leibler divergence}
\acrodef{sde}[SDE]{Shannon differential entropy}
\acrodef{ode}[ODE]{ordinary differential equations}
\acrodef{vb}[VB]{variational Bayes}
\acrodef{pmf}[PMF]{point-mass filter}
\acrodef{gmf}[GMF]{Gaussian mixture filter}
\acrodef{gm}[GM]{Gaussian mixture}
\acrodef{gf}[GF]{Gaussian filter}
\acrodef{gaf}[GAF]{Gaussian-assumed filter}
\acrodef{gpb}[GPB]{generalized pseudo-Bayesian}
\acrodef{imm}[IMM]{interacting multiple-model}
\acrodef{ut}[UT]{unscented transform}
\acrodef{ekf}[EKF]{extended \ac{kf}}
\acrodef{ckf}[CKF]{cubature \ac{kf}}
\acrodef{ukf}[UKF]{unscented \ac{kf}}
\acrodef{kf}[KF]{Kalman filter} 
\acrodef{pf}[PF]{particle filter} 
\acrodef{sif}[SIF]{stochastic integration filter} 
\acrodef{bpf}[BPF]{bootstrap \ac{pf}} 
\acrodef{kg}[KG]{Kalman gain}
\acrodef{ai}[AI]{artificial intelligence} 
\acrodef{dnn}[DNN]{deep neural network} 
\acrodef{cnn}[CNN]{convolutional neural network} 
\acrodef{nn}[NN]{neural network}  
\acrodef{mlp}[MLP]{multilayer perceptron}  
\acrodef{gnn}[GNN]{graph neural network} 
\acrodef{rnn}[RNN]{recurrent neural network} 
\acrodef{fc}[FC]{fully connected} 
\acrodef{snr}[SNR]{signal-to-noise ratio}
\acrodef{ml}[ML]{machine learning}
\acrodef{gru}[GRU]{gated recurrent unit} 
\acrodef{lstm}[LSTM]{long short-term memory} 
\acrodef{rkn}[RKN]{recurrent Kalman network}
\acrodef{pbm}[PBM]{physics-based model}
\acrodef{apbm}[APBM]{data-augmented physics-based model}
\acrodef{tm}[TM]{true model}
\acrodef{rts}[RTS]{Rauch-Tung-Striebel}
\acrodef{ddm}[DDM]{data-driven model}
\acrodef{vae}[VAE]{variational autoencoder}
\acrodef{pinn}[PINN]{physics-informed neural network}
\acrodef{rl}[RL]{reinforcement learning}
\acrodef{mcc}[MCC]{maximum correntropy criterion}
\acrodef{mdp}[MDP]{Markov decision problem}
\acrodef{pomdp}[POMDP]{partially observable Markov decision problem}
\acrodef{td}[TD]{temporal-difference}
\acrodef{narma}[NARMA]{nonlinear autoregressive moving-average}
\acrodef{armax}[ARMAX]{autoregressive moving-average model}
\acrodef{ungm}[UNGM]{univariate nonstationary growth model}
\acrodef{ctm}[CTM]{coordinated turn model}

\acrodef{nat}[NaT]{navigation and tracking}

\IEEEoverridecommandlockouts

\newcommand\clearrow{\global\let\rowmac\relax}
\clearrow

\usepackage[ruled,vlined]{algorithm2e}

\SetNlSty{mynlfont}{}{} 
\DontPrintSemicolon
\SetAlFnt{\footnotesize}

\usepackage{cite} 

\def\rmse{\mathrm{RMSE}}
\def\anees{\mathrm{ANEES}}
\def\mse{\text{MSE}}
\def\nis{\text{NIS}}

\def\px{\mathsf{x}}
\def\py{\mathsf{y}}
\def\vx{\dot{\mathsf{x}}}
\def\vy{\dot{\mathsf{y}}}

\begin{document}

\title{Adaptive Parameter Selection in Nonlinear State Estimation via Sequential Decision Making}
\title{Learning Adaptive Parameter Policies for Nonlinear Bayesian Filtering}

\author{%
	\IEEEauthorblockN{Ond\v{r}ej Straka}
	\IEEEauthorblockA{%
		European Centre of Excellence NTIS, \\
		University of West Bohemia in Pilsen, Czech Republic\\ E-mail: straka30@kky.zcu.cz}
	\and
	\IEEEauthorblockN{Felipe Giraldo-Grueso, and Renato Zanetti}
	\IEEEauthorblockA{
		Department of Aerospace Engineering and Engineering Mechanics\\
		The University of Texas at Austin, Austin, Texas 78712\\
		Email: \{fgiraldo, renato\}@utexas.edu}
	\thanks{ The work was partially supported by the Ministry of Education, Youth and Sports of the Czech Republic under project ROBOPROX - Robotics and Advanced Industrial Production CZ.02.01.01/00/22\_008/0004590.}
}%
\selectlanguage{english}
\maketitle
\begin{abstract}
For many nonlinear Bayesian state estimation problems, the posterior recursion is not analytically tractable, leading to algorithms that are influenced by numerical approximation errors. These algorithms depend on parameters that affect the approximation's accuracy and computational cost.
The parameters include, for example, the number of particles, scaling parameters, and the number of iterations in iterative computations. 
Typically, these parameters are fixed or adjusted heuristically, although the approximation accuracy can change over time with the local degree of nonlinearity and uncertainty.
The approximation errors introduced at a time step propagate through subsequent updates, affecting the accuracy, consistency, and robustness of future estimates. 
This paper presents adaptive parameter selection in nonlinear Bayesian filtering as a sequential decision-making problem, where parameters influence not only the immediate estimation outcome but also the future estimates.
The decision-making problem is addressed using reinforcement learning to learn adaptive parameter policies for nonlinear Bayesian filters.
Experiments with the unscented Kalman filter and stochastic integration filter demonstrate that the learned policies improve both estimate quality and consistency.
\end{abstract}

\section{Introduction}\label{sec:introduction}
State estimation for stochastic dynamical systems is traditionally addressed using the Bayesian framework~\cite{Sa:13}, leading to \acp{brr} that enable the calculation of the posterior \ac{pdf} of the state conditioned on previous available measurements.
For linear systems with Gaussian noise, the \ac{kf} provides an analytic recursive solution.
Such an analytic solution is available only for a few special cases, usually involving linear systems.
For nonlinear systems, a wide range of approximations have been developed, including the \ac{ekf}, \ac{ukf}~\cite{JuUhl:04}, \ac{sif}~\cite{DuStrSi:13}, ensemble-based filters~\cite{evensen1994sequential,ref:anderson,Yun2022_EnGMF}, Gaussian sum filters~\cite{Straka2025Efficient}, and \acp{pf}~\cite{DoFrGo-book:01}.

Despite their success, these methods share a common structural property: the state posterior \ac{pdf} update rule is fixed.
Given a previous state posterior~\ac{pdf} and a new observation, the estimator applies a predetermined update operator derived from local approximations or sampling schemes.
The update operator is often parametrized by a set of scalar parameters, sampling distribution~\cite{SiStra:03b}, or grid parameters~\cite{MaDuBr:23}.
Their specification is either fixed in time or adapted based on current measurements~\cite{StDuSi:14a,Dunik2012Unscented,Havlik2016OnNonlinearity,Popov2024_adaptive} to achieve optimal quality or the smallest possible approximation error with respect to the true posterior \ac{pdf}.
Optimality guarantees, where available, rely on restrictive assumptions such as linearity, Gaussianity, or correct model specification.
In practice, these assumptions are often violated.
Nonlinear measurement functions, heavy-tailed noise, and model mismatch often result in estimator inconsistency, divergence, or irreversible information loss, such as the premature collapse of multimodal posteriors.

In recent years, \ac{ml} techniques, particularly \ac{rl}, have been applied to improve the performance of classical estimators.
Representative approaches include \ac{ml}-based adaptation of filter parameters~\cite{Fan2024RNN-UKF,Scardua2017Complete}, \ac{rl}-based adaptation of process and measurement noise covariance matrices~\cite{Bekhtaoui2017Maneuvering,Shaaban2024Q-Learning}, direct learning of Kalman gains or correction terms~\cite{Hu2021LyapunovBased},	neural augmentation of prediction or update steps~\cite{Revach2021Kalmannet}, or end-to-end neural filters trained to output state estimates~\cite{Ghosh2024DANSE}.

These methods have demonstrated empirical improvements in some scenarios, especially under unknown or time-varying noise statistics.
However, they retain the underlying estimator structure: the posterior \ac{pdf} update mechanism remains fixed, and learning is confined to parameter tuning within that structure.
These approaches are unable to express non-myopic estimation strategies and cannot explicitly reason about the long-term consequences of posterior probability distribution updates.
Improvements are typically measured in terms of instantaneous or short-horizon error metrics such as mean squared error or its robust counterpart correntropy~\cite{Chen2016GeneralizedCorrentropy}, rather than long-term estimator reliability or stability.

In control theory, dynamic programming~\cite{Be:07a}, and \ac{rl}~\cite{Puterman2014Markov} are standard tools for handling long-horizon decision-making under uncertainty~\cite{Straka2022Distributed}.
In contrast, estimation is almost exclusively treated as an inference problem, despite its inherently sequential nature.
A small body of theoretical work has explored connections between Bayesian filtering and optimal control in posterior space, but these insights have rarely been exploited to design practical estimation algorithms~\cite{Strens2000Bayesian}.
In particular, there is no general framework that formulates posterior ~\ac{pdf} updates as decisions, optimizes estimator behavior over long horizons, and explains classical filters as special cases of optimal policies.

This paper aims to take a first step toward formulating the estimator as a decision maker. While keeping the filter update rule fixed, the filter parameters are optimized non-myopically to improve filter consistency. The problem is first formulated for an arbitrary parameter of any nonlinear Bayesian filter. Then, it is tailored to the adaptation of the \ac{ukf} scaling parameter and the number of \ac{sif} iterations.

Switching or jump-Markov filters, including interacting multiple model-type approaches, are related in that a discrete mode can be interpreted as a changing model parameter. In contrast, the present work adapts numerical parameters of a fixed approximate filter, such as the \ac{ukf} scaling parameter or the \ac{sif} iteration count, rather than inferring a latent switching mode in the generative model.

The paper is structured as follows: Section~\ref{sec:state_estimation} provides the formulation of the state estimation problem, introduces the Bayesian and optimization approaches, and discusses approximate solutions.
Section~\ref{sec:state_estimation_as_MDP} then formulates the Bayesian state estimation as a \ac{mdp} and shows its myopic property.
Section~\ref{sec:parameter_adaptation_as_MDP} then shows that parameter adaptation can be formulated as a non-myopic~\ac{mdp} and presents its solution using the \ac{rl} framework.
Application of the \ac{rl} framework to parameter adaptation in the Gaussian assumed filter is shown in Section~\ref{sec:algorithm}.
Numerical illustration is presented in Section~\ref{sec:numerical_illustration} and concluding remarks are drawn in Section~\ref{sec:conclusion}.
\section{State estimation problem}\label{sec:state_estimation}
\subsection{System model}
Consider a discrete-time stochastic system described by a nonlinear state-space model
\begin{subequations}\label{eq:sseq}
	\begin{align}\label{eq:sseqx}
		\bfx_{k+1} & = \bff_k(\bfx_k) + \bfw_k,                 \\
		\bfz_k     & = \bfh_k(\bfx_k) + \bfv_k,\label{eq:sseqz}
	\end{align}
\end{subequations}
where $\bfx_k\in\real^{n_x}$ and $\bfz_k\in\real^{n_z}$ represent the immeasurable state of the system and the available measurement at time instant $k=0,1,\ldots$, respectively. 
The functions $\bff_k:\real^{n_x}\mapsto\real^{n_x}$ and $\bfh_k:\real^{n_x}\mapsto\real^{n_z}$ are assumed known. The state noise $\bfw_k\in\real^{n_x}$ and measurement noise $\bfv_k\in\real^{n_z}$ are described by known \acp{pdf} $p_{\bfw_k}$ and $p_{\bfv_k}$, respectively. 
The initial state $\bfx_0$ is given by known \ac{pdf} $p_{\bfx_0}$.
Both noises are assumed to be white, mutually independent, and independent of $\bfx_0$.

The model~\eqref{eq:sseq} can be expressed using the transition \ac{pdf}~\eqref{eq:sspdfx} and measurement \ac{pdf}~\eqref{eq:sspdfz}
\begin{subequations}\label{eq:sspdf}
	\begin{align}\label{eq:sspdfx}
		p(\bfx_{k+1}|\bfx_k) & =p_{\bfw_k}(\bfx_{k+1}-\bff_k(\bfx_k)),                 \\
		p(\bfz_{k}|\bfx_k)   & =p_{\bfv_k}(\bfz_{k}-\bfh_k(\bfx_k)).\label{eq:sspdfz}
	\end{align}
\end{subequations}
\subsection{Bayesian state estimation}
The goal of state estimation in the Bayesian framework is to infer the posterior \ac{pdf} $p(\bfx_k|\bfz^k)$ of the state $\bfx_k$ given all the measurements available up to time~$k$, denoted as $\bfz^k\coloneqq[\bfz_1\T,\bfz_2\T,\cdots,\bfz_k\T]\T$.
The posterior \ac{pdf} is computed by the \acp{brr} consisting of the Bayes equation~\eqref{eq:brrbayes} and the \ac{cke}~\eqref{eq:brrcke}
\begin{subequations}\label{eq:brr}
	\begin{align}\label{eq:brrbayes}
		p(\bfx_k|\bfz^k)     & =\frac{p(\bfz_k|\bfx_k)p(\bfx_k|\bfz^{k-1})}{\int p(\bfz_k|\bfx_k)p(\bfx_k|\bfz^{k-1})\d\bfx_k}               \\
		p(\bfx_k|\bfz^{k-1}) & =\int p(\bfx_k|\bfx_{k-1})p(\bfx_{k-1}|\bfz^{k-1})\d\bfx_{k-1}.\label{eq:brrcke}
	\end{align}
\end{subequations}
The initial condition for the \acp{brr} is $p(\bfx_0|\bfz^0)=p(\bfx_0)$.
The calculation of the \acp{brr} thus involves alternating the filtering step~\eqref{eq:brrbayes} and the prediction step~\eqref{eq:brrcke}.
Usually, an approximate solution has to be used to obtain the filtering \ac{pdf} $p(\bfx_k|\bfz^k)$ and the predictive \ac{pdf} $p(\bfx_k|\bfz^{k-1})$.
\subsection{State estimation as an optimization problem}
State estimation can also be formulated as a search for a point estimate of $\bfx_k$ given the available information $\bfz^k$. The estimate is denoted as $\hbfx_k(\bfz^k)$. If the criterion $J(\tbfx_k)$ to be minimized with $\tbfx_k(\bfz^k)\coloneqq\bfx_k-\hbfx_k(\bfz^k)$ representing the estimate error is the \ac{mse}
\begin{align}\label{eq:msecrit}
	J_{\mse}(\tbfx_k)=\mean\left[(\tbfx_k)\T\tbfx_k\right],
\end{align}
the optimum estimate is the conditional mean $\hbfx(\bfz^k)=\mean[\bfx_k|\bfz^k]$. Replacing the squared L2-norm in~\eqref{eq:msecrit} with the L1-norm yields the median. Additional alternatives are the correntropy criterion~\cite{Chen2016GeneralizedCorrentropy} and the Huber cost function~\cite{Karlgaard2007Huber}.

\subsection{Approximate solutions}
An analytic solution to~\eqref{eq:brr} can be obtained only for a few special combinations of the model~\eqref{eq:sseq} and noise \acp{pdf}. 
Mostly, an approximate solution must be searched for. Assuming the Gaussian joint \ac{pdf} $p(\bfz_k,\bfx_k|\bfz^{k-1})$ in the numerator of~\eqref{eq:brrbayes} leads to the Gaussian-assumed filters such as the \ac{ckf}~\cite{Sa:13} or \ac{sif}~\cite{DuStrSi:13}. 
Assuming a Gaussian mixture distributed joint \ac{pdf} leads to the Gaussian mixture filter~\cite{Straka2025Efficient} and assuming the joint \ac{pdf} being Student-t leads to the Student-t filter~\cite{Straka2017Stochastic}. 
Since the paper demonstrates parameter adaptation using Gaussian-assumed filters, its generic algorithm is described in Algorithm~\ref{alg:GF} for the zero-mean Gaussian noises $\bfw_k$ and $\bfv_k$ with covariance matrices $\bfQ_k$ and $\bfR_k$, respectively, and Gaussian initial condition $p(\bfx_0)=\calN\{\bfx_0;\bbfx_0,\bfP_0\}$.
The algorithm calculates the mean and covariance matrix of the prediction \ac{pdf} $p(\bfx_{k+1}|\bfz^k)=\calN\{\bfx_{k+1};\hbfx_{k+1|k},\bfP_{k+1|k}^{\bfx\bfx}\}$ and filtering \ac{pdf} $p(\bfx_{k+1}|\bfz^{k+1})=\calN\{\bfx_{k+1};\hbfx_{k+1|k+1},\bfP_{k+1|k+1}^{\bfx\bfx}\}$

\begin{algorithm}[t]
	\caption{Gaussian assumed filter}
	\label{alg:GF}
	\DontPrintSemicolon

	\textbf{Initialization:}\;
	Set $k \leftarrow 0$\;
	Define the initial condition $p(\bfx_0|\bfz^{0}) =  \calN\{\bfx_0;\hbfx_{0|0}, \bfP^{\bfx\bfx}_{0|0}\}$,
	with $\hbfx_{0|0}=\bbfx_0$ and $\bfP^{\bfx\bfx}_{0|0}=\bfP_0$\;
	\While{new measurement $\bfz_{k+1}$  available}{
		\textbf{Predict:}\;
		Compute the state prediction mean and covariance:
		\begin{align*}
			\hbfx_{k+1|k}           & = \mean[\bff_k(\bfx_{k})|\bfz^k]                          \\
			\bfP^{\bfx\bfx}_{k+1|k} & = \mean[(\bff_k(\bfx_{k})-\hbfx_{k+1|k})(\cdot)\T|\bfz^k]
			+ \bfQ_k
		\end{align*}
		\textbf{Update:}\;
		Compute the measurement prediction mean and covariance:
		\begin{align}
			\label{eq:GFzp}
			\hbfz_{k+1|k}           & = \mean[\bfh_{k+1}(\bfx_{k+1})|\bfz^{k}]                                                    \\
			\label{eq:GFPzz}
			\bfP^{\bfz\bfz}_{k+1|k} & = \mean[(\bfh_{k+1}(\bfx_{k+1})-\hbfz_{k+1|k})(\cdot)\T|\bfz^{k}] + \bfR_{k+1}              \\
			\label{eq:GFPxz}
			\bfP^{\bfx\bfz}_{k+1|k} & = \mean[(\bfx_{k+1}{-}\hbfx_{k+1|k}) (\bfh_{k+1}(\bfx_{k+1}){-}\hbfz_{k+1|k})\T|\bfz^{k}]
		\end{align}

		Compute the gain $\bfK_{k+1} = \bfP^{\bfx\bfz}_{k+1|k} (\bfP^{\bfz\bfz}_{k+1|k})^{-1}$

		Update mean using $\tbfz_{k+1|k}\coloneqq(\bfz_{k+1}-\hbfz_{k+1|k})$ as
		\begin{align*}
			\hbfx_{k+1|k+1} &= \hbfx_{k+1|k} + \bfK_{k+1}\tbfz_{k+1|k} 
		\end{align*}

		Update covariance:
		\begin{align*}
			\bfP^{\bfx\bfx}_{k+1|k+1} = \bfP^{\bfx\bfx}_{k+1|k} - \bfK_{k+1} \bfP^{\bfz\bfz}_{k+1|k} \bfK_{k+1}\T
		\end{align*}

		$k \leftarrow k+1$\;
	}

	\textit{Note:} $(\bfa)(\cdot)\T$ denotes $(\bfa)(\bfa)\T$.\;

\end{algorithm}

\section{State estimation as a Markov decision problem}\label{sec:state_estimation_as_MDP}
When formulating the state estimation as a sequential optimization problem, for all time instants $k\geq0$, we aim to derive an estimator $\hbfx_k(\bfz^k)$ based on the available information.
To do this, we introduce a loss function $L(\bfx_k,\hbfx_k(\bfx^k))$ and seek to minimize the cumulative expected loss defined as follows
\begin{align}\label{eq:cost}
	J = \lim_{T\rightarrow \infty}\mean_{\bfx^T,\bfz^T}\left[\sum_{k=0}^{T}\gamma^k L(\bfx_k,\hbfx_k(\bfz^k))\right],
\end{align}
where $\gamma\in(0,1]$ is the discount factor.

Since the state $\bfx_k$ is not available, a reformulation is introduced, replacing the unknown state $\bfx_k$ by an information state $\bfs_k$~\cite{Astrom1965Optimal}. 
The information state $\bfs_k$ is defined as the conditional \ac{pdf}
\begin{align}\label{eq:istate}
	\bfs_k\coloneqq p(\bfx_k|\bfz^k).
\end{align}
The information state $\bfs_k$ dynamics is given by
\begin{align}\label{eq:istateupdate}
	\bfs_{k+1}=\bfPhi_\text{BRR}(\bfs_k,\bfz_{k+1}),
\end{align}
where $\bfPhi_\text{BRR}$ is the Bayesian update operator composed of~\eqref{eq:brrbayes} and~\eqref{eq:brrcke}.
From~\eqref{eq:istateupdate}, it follows that the information state process is Markov. Due to the random nature of the measurement, the transition of the information state is stochastic.

Using the information state, the sequential optimization problem with the cumulative loss~\eqref{eq:cost} can be reformulated as
\begin{align}\label{eq:cost_is}
	J = \lim_{T\rightarrow \infty}\mean_{\bfz^T}\left[\sum_{k=0}^{T}\gamma^k \bL(\bfs_k,\hbfx_k)\right],
\end{align}
where
\begin{align}
	\bL(\bfs_k,\hbfx_k)=\mean_{\bfx_k|\bfz^k}[L(\bfx_k,\hbfx_k)]=\int_{\real^{n_x}} L(\bfx_k,\hbfx_k)\bfs_k(\bfx_k)\d\bfx_k.
\end{align}
The solution $\hbfx_k(\bfz^k)$ to the original problem~\eqref{eq:cost} now becomes $\hbfx_k(\bfs_k)$.
The solution to the reformulated problem is identical to that of the original problem~\cite{Astrom1965Optimal}. This problem is called \ac{mdp}, more specifically, the \ac{pomdp} due to the unavailability of $\bfx_k$.

For this reformulated problem, the Bellman recursion~\cite{Be:07a} has the form
\begin{align}\label{eq:bellman_v}
	V(\bfs_k) = \min_{\hbfx_k}\left[ \bL(\bfs_k,\hbfx_k) + \gamma \mean\left[V\left(\bfPhi_\text{BRR}(\bfs_k,\bfz_{k+1})\right)\right]\right],
\end{align}
where $V:\calS\rightarrow\real$ is the Bellman (value) function.
The optimum estimate is then
\begin{align}\label{eq:bellman_x}
	\hbfx_k(\bfs_k) = \arg\min_{\hbfx_k}\left[ \bL(\bfs_k,\hbfx_k) + \gamma \mean\left[V\left(\bfPhi_\text{BRR}(\bfs_k,\bfz_{k+1})\right)\right]\right].
\end{align}
Since the second term in the RHS of~\eqref{eq:bellman_x} does not depend on $\hbfx_k$, the estimator becomes myopic
\begin{align}
	\hbfx_k(\bfs_k) = \argmin \bL(\bfs_k,\hbfx_k)
\end{align}
and no multi-step coupling exists.
The optimum estimate then corresponds to that computed using the standard non-\ac{mdp} optimization procedure.

\section{Parameter Adaptation as \ac{mdp}}\label{sec:parameter_adaptation_as_MDP}
For the nonlinear or non-Gaussian models, the Bayesian update~\eqref{eq:brr} is typically intractable in closed form, motivating an approximate posterior representation.
The approximate posterior generated by an estimator is typically characterized by a set of parameters $\bftheta\in\Theta$ utilized by that estimator.
The quality of the approximation at time $k$ then depends on previous decisions (values of $\bftheta_t$, $t\leq k$), and similarly, the current parameters $\bftheta_k$ influence future approximations.
The problem of parameter adaptation will now be formulated as an \ac{mdp}, as in the previous section.
However, instead of looking for an estimate $\hbfx_k$, we will look for parameters $\bftheta_k$ assuming a fixed structure of the estimator (e.g., the Gaussian-assumed filter such as the \ac{ukf}). 

\subsection{Specification of information state, parameters, and cost}
As the posterior \ac{pdf} is not available, the information state is assembled from
the estimator's internal representation of the state \ac{pdf}, such as the predictive state mean and covariance, sigma points, or a set of particles. The new information state will be denoted as $\bbfs_k\in\bar{\calS}$ to distinguish it from $\bfs_k$.
Then, the information state $\bbfs_k$ dynamics can be written as\footnote{Note that, in contrast to~\eqref{eq:istateupdate}, the information state $\bbfs_{k+1}$ depends on $\bfz_k$ as $\bbfs_{k+1}$ is a predictive estimate while $\bfs_{k+1}$ represented a filtering estimate.}
\begin{align}\label{eq:istateupdateaction}
	\bbfs_{k+1}=\bfPhi_\text{filter}(\bbfs_k,\bfz_{k},\bftheta_{k}).
\end{align}
In the \ac{mdp} context, the estimator's parameters now serve as actions that affect the future information state.
They are generated by a \emph{policy} $\pi: \bar{\calS}\times\real^{n_z} \rightarrow \Theta$ that assigns a parameter $\bftheta_k$ to each decision situation at time $k$ characterized by the information state and current measurement:
\begin{align}
\bftheta_k \sim \pi(\bbfs_k,\bfz_k).
\end{align}
The parameters~$\bftheta_k$ may include the number of \ac{pf} samples, the model selection criteria, or the \ac{ukf} scaling parameter.

The parameters $\bftheta_k$ enter the cost $\bL$ through the estimate $\hbfx_k(\bftheta_k)$  $\bL(\bbfs_k,\hbfx_k(\bftheta_k))$.
The specification of such a cost may be tricky, as the reference value for the estimate $\hbfx_x(\bftheta_k)$ is calculated by the same estimator. 
A more convenient form of the cost function is defined in terms of the measurement-related quantities such as the measurement $\bfz_k$, its prediction depending on the parameters $\hbfz_{k|k-1}(\bftheta_k)$, and the covariance $\bfP_{k|k-1}^{\bfz\bfz}(\bftheta_k)$.
 Such cost can be denoted by $\bL(\hbfz_{k|k-1}(\bftheta_k),\bfP_{k|k-1}^{\bfz\bfz}(\bftheta_k),\bfz_k)$.
However, for the sake of convenience, the notation $\bL(\bbfs_k,\bfz_k,\bftheta_k)$ expressing the dependence of the cost on the information state, the measurement, and parameters will be used in the sequel.

In general, the cost $\bL$ may value estimate \emph{quality} given by a norm of measurement prediction error $\bL(\bbfs_k,\bfz_k,\bftheta_k) = \|\tbfz_{k|k-1}(\bftheta_k)\|$
or \emph{consistency} given by \ac{nis} $\bL(\bbfs_k,\bfz_k,\bftheta_k) = \tbfz_{k|k-1}(\bftheta_k)\T(\bfP_{k|k-1}^{\bfz\bfz}(\bftheta_k))^{-1}
    \tbfz_{k|k-1}(\bftheta_k)$
 or \emph{robustness} given by penalizing large estimate errors, computational costs, or by a combination thereof. 

With the above considerations, the Bellman recursion for the parameter adaptation has the form \begin{align}\label{eq:bellman_par_v}
	V(\bbfs_k,\bfz_k) =& \min_{\bftheta_k}\big[ \bL(\bbfs_k,\bfz_k,\bftheta_k)\nonumber\\
    &+ \gamma \,\mean\left[V\left(\bfPhi_\text{filter}(\bbfs_k,\bfz_{k},\bftheta_{k}),\bfz_{k+1}\right)\right]\big].
\end{align}
The expectation on the right-hand side of~\eqref{eq:bellman_par_v} is due to the presence of the future measurement $\bfz_{k+1}$ and also due to the cases when the filter information state dynamics~\eqref{eq:istateupdateaction} contains random effects, which is the case of \acp{pf}.
The optimal adaptive parameter policy $\pi$ providing $\bftheta_k$ is then
\begin{align}\label{eq:parameter_adaptation}
	\pi(\bbfs_k,\bfz_k) =& \arg\min_{\bftheta_k}\big[ \bL(\bbfs_k,\bfz_k,\bftheta_k). \nonumber\\ + & \gamma\, \mean\left[V\left(\bfPhi_\text{filter}(\bbfs_k,\bfz_{k},\bftheta_{k}),\bfz_{k+1}\right)\right]\big].
\end{align}

By this (PO)\ac{mdp} formulation, the estimation problem becomes sequentially coupled, adaptive, and non-myopic. 
It accounts for the fact that approximation quality depends on earlier parameter choices.

In contrast, the classical approach to state estimation is inherently myopic. It focuses on obtaining the best possible estimate at each moment in time by calculating an approximate estimate that is closest to the optimal value, such as the conditional mean in the case of \ac{mse}. 
In the classical formulation, one cannot worsen the current estimate to achieve a substantial improvement in the future.
\subsection{Reinforcement learning solution to \ac{mdp}}
Analytical solution to~\eqref{eq:parameter_adaptation} is not tractable, and in this paper, an~\ac{ml} solution is chosen.
In particular, the~\ac{rl} approach is adopted, which falls under unsupervised \ac{ml}.
We employ the on-policy actor-critic algorithm with temporal-difference learning TD(0)~\cite{Sutton1998Reinforcement} with the following assumptions:
The parameter space $\Theta$ is assumed discrete $\bftheta\in\Theta=\{\bftheta^{(i)},i=1,\ldots |\Theta|\}$.

The adaptive parameter policy $\pi(\bbfs_k,\bfz_k)$ (actor) is assumed stochastic, given by a conditional probability distribution $\pi_{\bftheta}(\bftheta|\bbfs_k,\bfz_k)$ and represented by a~\ac{nn}
\begin{align}
	\pi_{\bftheta}(\bftheta|\bbfs_k,\bfz_k;\phi) = \text{softmax}(f_\phi(\bbfs_k,\bfz_k)),
\end{align}
where $f_\phi(\cdot)$ is a feedforward \ac{mlp} \ac{nn} parametrized by $\phi$.
The Bellman function $V$ (critic) is approximated by another \ac{mlp} \ac{nn}    $V_{\psi}(\bbfs_k,\bfz_k)$ parametrized by $\psi$.

Both actor and critic are updated based on the \ac{td} error computed as (c.f.~\eqref{eq:bellman_par_v})
\begin{align}\label{eq:tde}
	\delta_k = \bL(\bbfs_k,\bfz_k,\bftheta_k) + \gamma\,V_\psi(\bbfs_{k+1},\bfz_{k+1})-V(\bbfs_k,\bfz_k).
\end{align}
The \ac{td} error is the discrepancy between the current estimate $V(\bbfs_k,\bfz_k)$ and one-step bootstrap target $\bL(\bbfs_k,\bfz_k,\bftheta_k) + \gamma\,V_\psi(\bbfs_{k+1},\bfz_{k+1})$.
The critic then minimizes the squared error
\begin{align}\label{eq:critic_update}
	\calL_{\text{critic}} = \delta_k^2
\end{align}
to learn the Bellman function.
The actor is updated to minimize the policy gradient objective
\begin{align}\label{eq:actor_update}
	\calL_\text{actor} = -\delta_k \log \pi_{\bftheta}(\bftheta|\bbfs_k,\bfz_k;\phi).
\end{align}
To stabilize the \ac{td} learning process, a target critic network is used. 
Instead of bootstrapping from the current critic $V_{\psi}(\bbfs_{k+1},\bfz_{k+1})$, the \ac{td} target is computed using a lagged copy $V_{\bar{\psi}}(\bbfs_{k+1},\bfz_{k+1})$.
This prevents the critic from chasing a moving target, mitigates divergence due to function approximation, and reduces variance in the value update. 
The target parameters are updated via soft (Polyak) averaging~\cite{Adamczyk2025Average}.

\section{Algorithm for Parameter Adaptation}\label{sec:algorithm}
The idea of parameter adaptation for nonlinear Bayesian filters is general and can be applied to any algorithm, such as the \ac{pf}, for adaptation of the particle number or the proposal density, the Gaussian mixture filter~\cite{Straka2025Efficient} for the specification of the number of Gaussian components, or the Gaussian assumed filters such as the \ac{ukf} for the specification of the scaling parameter, or for the \ac{sif} for specification of the number of iterations.

The algorithm for offline learning for the Gaussian assumed filter is described in Algorithm~\ref{alg:offline_ac}. 
To simplify the exposition, only parameters appearing in the update step, i.e., the calculation of the predictive measurement moments (\ref{eq:GFzp}-\ref{eq:GFPxz}) are adapted.
The steps labeled~\ref{algl:predict} and~\ref{algl:update} correspond to the predict and update steps in Algorithm~\ref{alg:GF}.

\begin{algorithm}[t]
	\caption{Offline Monte-Carlo Actor--Critic Training for Adaptive Parameter Policy}
	\label{alg:offline_ac}

	\KwIn{Discrete parameter set $\Theta=\{\bftheta^{(1)},\dots,\bftheta^{(|\Theta|)}\}$,
		discount factor $\gamma$, number of episodes $N_{\text{MC}}$, horizon $T$, hyperparameter $\tau$, actor $\pi_\theta(\theta|\bbfs,\bfz;\phi)$, critic $V_{\psi}(\bbfs,\bfz)$}
	\KwOut{Trained actor parameters $\phi$ and critic parameters $\psi$}\;

	Initialize actor parameters $\phi$, critic parameters $\psi$, and target critic parameters $\bar{\psi} \leftarrow \psi$\;

	\For{$m = 1$ \KwTo $N_{\text{MC}}$}{
	Reset simulator and filter state $(\hbfx_{0|0}$, $\bfP_{0|0}^{\bfx\bfx})$\;

	\For{$k = 1$ \KwTo $T-1$}{
    
\lnl{algl:predict}
	$(\hbfx_{k|k-1}, \bfP_{k|k-1}^{\bfx\bfx}) \leftarrow \text{Predict}(\hbfx_{k-1|k-1}, \bfP_{k-1|k-1}^{\bfx\bfx})$\;

	Construct $\bbfs_{k}$ from $\hbfx_{k|k-1}$ and $\bfP_{k|k-1}^{\bfx\bfx}$\;

	Sample $\bftheta_{k} \sim \pi_{\bftheta}(\cdot|\bbfs_k,\bfz_k;\phi)$\;

\lnl{algl:update}
	$(\hbfx_{k|k}, \bfP_{k|k}^{\bfx\bfx}, \tbfz_{k|k-1}, \bfP_{k|k-1}^{\bfz\bfz})
		\leftarrow \text{Update}(\hbfx_{k|k-1}, \bfP_{k|k-1}^{\bfx\bfx}, \bftheta_k)$\;

Compute cost $L_k(\bbfs_k,\bfz_k,\bftheta_k)$\;

	$(\hbfx_{k+1|k}, \bfP_{k+1|k}^{\bfx\bfx})
		\leftarrow \text{Predict}(\hbfx_{k|k}, \bfP_{k|k}^{\bfx\bfx})$\;

    
	Construct $\bbfs_{k+1}$ from $\hbfx_{k+1|k}$ and $\bfP_{k+1|k}^{\bfx\bfx}$\;

		$\delta_k \leftarrow L_k(\bftheta_k,\bbfs_k,\bfz_k) + \gamma V_{\bar{\psi}}(\bbfs_{k+1},\bfz_{k+1}) - V_{\psi}(\bbfs_k,\bfz_k)$\;

	$\calL_{\text{critic}} \leftarrow \delta_k^2$\;

	$\calL_{\text{actor}} \leftarrow
		-\delta_k \log \pi_{\bftheta}(\bftheta_k | \bbfs_k,\bfz_k;\phi)$\;

	Update $\psi$ and $\phi$ using Adam optimizer\;

Update target critic parameters:
		$\bar{\psi} \leftarrow \tau \psi + (1-\tau)\bar{\psi}$\;
	}
	}
\end{algorithm}

After learning the policy $\pi_{\bftheta}(\bftheta|\bbfs_k,\bfz_k;\phi)$, the Gaussian assumed filter with parameter adaptation can be used. Its algorithm is described in Algorithm~\ref{alg:GFPA}.

\begin{algorithm}[t]
	\caption{Gaussian assumed filter with adaptive parameter policy}
	\label{alg:GFPA}
	\DontPrintSemicolon

	\KwIn{Adaptive parameter policy $\pi_\theta(\theta|\bbfs,\bfz;\phi)$}
    
	\textbf{Initialization:}\;
	Set $k \leftarrow 0$\;
	Define the initial condition
	\[
		p(\bfx_0|\bfz^{0}) = p(\bfx_0)
		= \calN\{\bfx_0;\hbfx_{0|0}, \bfP^{\bfx\bfx}_{0|0}\},
	\]
	with $\hbfx_{0|0}=\bbfx_0$ and $\bfP^{\bfx\bfx}_{0|0}=\bfP_0$\;

	\While{new measurement $\bfz_{k+1}$  available}{

		\textbf{Predict:}\;
		Compute the state prediction mean $\hbfx_{k+1|k}$ and covariance $\bfP^{\bfx\bfx}_{k+1|k}$ as in Algorithm~\ref{alg:GF}.
        
	Construct $\bbfs_{k+1}$ from $\hbfx_{k+1|k}$ and $\bfP_{k+1|k}^{\bfx\bfx}$\;

	Select $\bftheta_{k+1}^*=\argmax_{\bftheta}\pi_{\bftheta}(\bftheta|\bbfs_{k+1},\bfz_{k+1};\phi)$\;
		\textbf{Update:}\;
		Compute the measurement prediction mean $\hbfz_{k+1|k}$ and covariances $\bfP^{\bfz\bfz}_{k+1|k}$ and $\bfP^{\bfx\bfz}_{k+1|k}$ using the optimum parameter $\bftheta_{k+1}^*$.\;
        Subsequently, compute the gain $\bfK_{k+1}$ and update the mean $\hbfx_{k+1|k+1}$ and covariance $\bfP^{\bfx\bfx}_{k+1|k+1}$ as in Algorithm~\ref{alg:GF}.




		$k \leftarrow k+1$\;
	}
\end{algorithm}
Notice that during offline learning, the parameter is sampled from the policy, while during online estimation, the greedy parameter value with respect to the policy is selected.
\section{Numerical illustration}\label{sec:numerical_illustration}
The proposed learning of adaptive parameter policy is demonstrated for the \ac{ukf} scaling parameter and the \ac{sif} number of iterations. 
For the demonstration, two estimation problems will be considered: the \ac{ungm} used often for benchmarking due to its strong nonlinearity and multimodal posterior, and the \ac{ctm} with bearing measurements. Both models are specified first.
\subsection{Univariate Nonstationary Growth Model}
The \ac{ungm} is a scalar nonlinear state–space model
\begin{align}
x_k = f(x_{k-1},k{-}1) + w_{k-1},
\qquad
w_{k-1}\sim\calN\{0,Q\},
\end{align}
with  $f(x,k)=0.5x+\tfrac{25x}{1+x^2}+8\cos(0.05\,k)$ and measurement equation
\begin{align}
y_k =\tfrac{x_k^2}{20} +v_k,
\qquad
v_k\sim\calN\{0,R\}.
\end{align}
The model was simulated for $k=1,\ldots 500$ with Gaussian initial state $x_1\sim\calN\{0,\,5)$ and variances $Q=1$ and $R=0.1$.
\subsection{Coordinated Turn Model}
The \ac{ctm} considers a four-dimensional state $\mathbf{x}_k=[\px_k,\vx_k,\py_k,\vy_k]\T\in\mathbb{R}^4$ consisting of positions $[\px,\py]$ and velocities $[\vx,\vy]$ in 2D space. 
It evolves according to a linear dynamics with additive Gaussian noise
\begin{equation}
\bfx_k = \bfF\,\bfx_{k-1} + \bfw_{k-1},
\qquad
\bfw_{k-1}\sim\calN\{\bfnul,\bfQ\},
\end{equation}
where the transition matrix is
\begin{align}
\bfF =
\begin{bsmallmatrix}
1 & \frac{\sin(\omega\Delta t)}{\omega} & 0 & -\frac{1-\cos(\omega\Delta t)}{\omega}\\
0 & \cos(\omega\Delta t) & 0 & -\sin(\omega\Delta t)\\
0 & \frac{1-\cos(\omega\Delta t)}{\omega} & 1 & \frac{\sin(\omega\Delta t)}{\omega}\\
0 & \sin(\omega\Delta t) & 0 & \cos(\omega\Delta t)
\end{bsmallmatrix}
\end{align}
with $\Delta t=1~\si{s}$ denoting the sampling period and $\omega$ being a known constant turn rate.
The process noise covariance is
\begin{align}
\mathbf{Q}
=
q\,
\Big(
\bfI_2 \otimes
\begin{bsmallmatrix}
\frac{\Delta t^3}{3} & \frac{\Delta t^2}{2}\\
\frac{\Delta t^2}{2} & \Delta t
\end{bsmallmatrix}
\Big),
\end{align}
with $\bfI_2$ being the identity matrix and $q$ being noise intensity.
The position is measured via bearing and is given by
\begin{align}
y_k =\operatorname{atan2}\!\big(\py_k,\px_k\big) + v_k,
\qquad
v_k\sim\calN\{0,R\}.
\end{align}
 Note that the innovation $\tbfz_{k|k-1}$ in the filter is wrapped to the interval $(-\pi,\pi]$.

The model was simulated for $k=1,\ldots 150$ with $\Delta t=1$, $\omega=0.5$, intensity $q=1$, variance $R=0.04$, and Gaussian  initial state
\begin{align}
\bfx_1 \sim \calN\{[80,\,0,\,0\,20]\T,\diag([10^3,\,10^2,\,10^3,\,10^2])\},
\end{align}
where $\diag$ denotes a diagonal matrix.
\subsection{Actor critic \ac{rl} parameters}
The actor was a feed-forward \ac{nn} with two 64-unit ReLU hidden layers mapping $\bbfs_k$ and $\bfz_k$ to a softmax distribution over discrete parameter choices.
The critic was a feed-forward \ac{nn} with two 64-unit ReLU hidden layers.
The actor and critic were optimized with Adam using learning rates of $10^{-4}$ and $5\times 10^{-4}$, respectively, with gradient decay $0.9$ and squared-gradient decay $0.999$.
A target critic with soft updates ($\tau=0.01$) was employed for stabilization, and entropy regularization with coefficient  $10^{-3}$ was used to promote exploration.
The forgetting factor was set to $\gamma=0.5$.
The \acp{nn} were trained using $N_\text{MC}=10^3$ simulations.
\subsection{Algorithms}
The learning of adaptive parameter policies is demonstrated for the adaptation of \emph{(i)}~the scaling parameter $\kappa$ of the \ac{ukf} and \emph{(ii)}~the number of iterations $N_\text{it}$ used in the \ac{sif}.
In the case of the \ac{ukf}, the computational cost was not included in the criterion since all choices of $\kappa$ are equivalent in terms of computational complexity.
On the other hand, the number of iterations used by the \ac{sif} to compute the predictive measurement moments $\hbfz_{k+1|k}$, $\bfP^{\bfz\bfz}_{k+1|k}$, $\bfP^{\bfx\bfz}_{k+1|k}$ has a significant impact on the computational complexity, which was, thus, included in the optimized cost.

The online overhead of the proposed adaptive policy is one forward evaluation of the actor network and the selection of the most probable discrete parameter value. This cost is typically small compared with the cost of evaluating sigma-point or stochastic-integration moments, especially for the \ac{sif}, where the number of iterations directly scales the update cost.

The information state $\bbfs_k$ for both cases consisted of the predictive moments computed by the algorithm, the measurement\footnote{Remind that the measurement $\bfz_k$ is an argument of the Bellman function~$V$.}, and the innovation $\tbfz_{k|k-1}$ and was defined as
\begin{align}
    \bbfs_k = [\hbfx_{k|k-1}\T,\, \tr(\bfP_{k|k-1}^{\bfx\bfx}),\, \log\det(\bfP_{k|k-1}^{\bfx\bfx}),\, \bfz_k\T,\, \tbfz_{k|k-1}\T]\T.
\end{align}

Several baseline settings were considered to analyze the performance of the adaptive parameter policy. 
\begin{itemize}
    \item  First, a \textbf{default} baseline employing the parameter value typically specified in the target application, representing standard practice where the filter is deployed with a fixed, heuristically chosen tuning. For the \ac{ukf}, it is $\kappa=\max(0,3-n_x)$, for the \ac{sif}, it is $N_\text{it}=10$.
    \item  Second, a set of \textbf{fixed} parameter baselines across various constant values providing a controlled overview of performance, highlighting the sensitivity of estimation accuracy and consistency to tuning choices.
    \item Third, a \textbf{myopic} baseline applying the same optimization criterion as the adaptive method but restricting the decision to a single-step horizon ($\gamma=0$), thereby isolating the effect of non-myopic reasoning and long-term trade-offs. 
    \item Fourth, an \textbf{optimal} baseline obtained by selecting the parameter that maximizes the likelihood~\cite{DuSiStra:10a}. Note that this baseline applies only to \ac{ukf} since for the \ac{sif}, the optimal choice is always the maximum number of iterations that is possible.
\end{itemize}
Collectively, these baselines establish a range of reference operating points, from conventional heuristic tuning to optimal values, enabling a systematic assessment of the benefits of adaptive, non-myopic parameter adaptation.
Note that the fixed and default baselines used fixed parameter values for all time instants, whereas the myopic and optimal baselines, and the proposed parameter adaptation, use parameters that vary over time according to their respective criteria.

The adaptation considered the following costs:
\begin{itemize}
    \item \ac{nis} cost penalizing consistency 
\begin{align}\label{eq:LNIS}
    \bL_\text{NIS} =& \left(\nis-n_z\right)^2,
\end{align}
where $\nis=\tbfz_{k|k-1}\T(\bfP_{k|k-1}^{\bfz\bfz})^{-1}\tbfz_{k|k-1}$ (for consistent measurement prediction estimates, $\mean[\nis]=n_z$)
    \item a \ac{nis}-based cost penalizing only optimistic estimates 
\begin{align}\label{eq:LlogmaxNIS}
    \bL_\text{logmaxNIS} =& \log\left(1+\max\left(0,\left(\nis-n_z\right)\right)\right),
\end{align}
\item a quality-oriented cost as an $L_2$ norm of the state innovation
\begin{align}\label{eq:stateinnov}
    \bL_\text{stateInnov}=&\|\bfK_k\tbfz_{k|k-1}\|.
\end{align}

\end{itemize}
\subsection{ Performance criteria}
The performance of the proposed parameter adaptation and baselines was measured using the \ac{rmse} defined as
\begin{align}
    \rmse_k = \sqrt{\frac{1}{M}\sum_{\ell=1}^{M}\tbfx_{k|k}(\ell)\T\,\tbfx_{k|k}(\ell)}
\end{align}
based on $M=10^4$ \ac{mc} simulations with $\tbfx_{k|k}(\ell)\coloneq \bfx_k(\ell)-\hbfx_{k|k}(\ell)$ being the estimate error,  $\bfx_k(\ell)$ being the true state at $\ell$-th \ac{mc} simulation and $\hbfx_{k|k}(\ell)$ being its filtering estimate. 
To assess higher-order information provided by the filters, the \ac{anees}~\cite{LiZha:06a} defined by
\begin{align}
    \anees_k = \frac{1}{M}\sum_{\ell=1}^{M}\tbfx_{k|k}(\ell)\T\,\cov[\bfx_k|\bfz^k]^{-1}\,    \tbfx_{k|k}(\ell)
\end{align}
is used. 
\ac{anees} assesses the consistency of the estimator, i.e., alignment of the conditional covariance matrix $\cov[\bfx_k|\bfz^k]$ and the estimate error $\tbfx_{k|k}$. 
This value should be close to $n_x$. 
Higher \ac{anees} values mean that the estimator is too optimistic, while values smaller than one mean the estimator is too pessimistic.

Note that the values of the costs $\bL$ described above are only loosely connected to \ac{anees}. 
The costs are defined in measurement space and use the actual value of the measurement as a reference value and its prediction.
The \ac{anees}, on the other hand, is defined in the state space and uses the true value of the state and the filtering estimate.

\subsection{Results}
\begin{figure}
    \centering
    \includegraphics[width=0.7\linewidth]{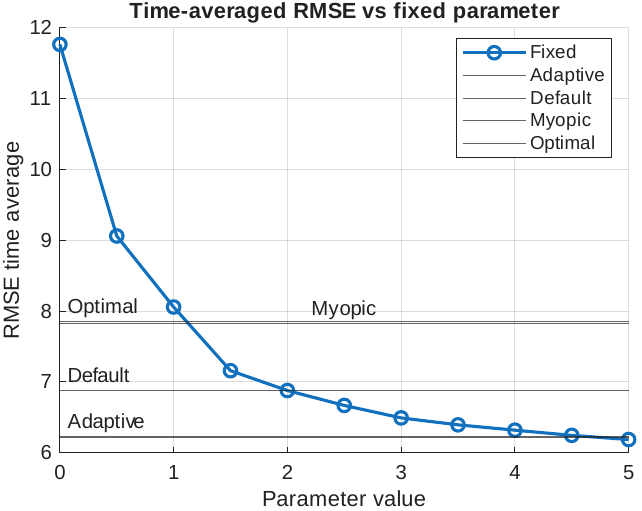}
    \caption{Time averaged \ac{rmse} for \ac{ukf} and \ac{ungm}.}
    \label{fig:ungm_ukf_rmse}
\end{figure}

\begin{figure}
    \centering
    \includegraphics[width=0.7\linewidth]{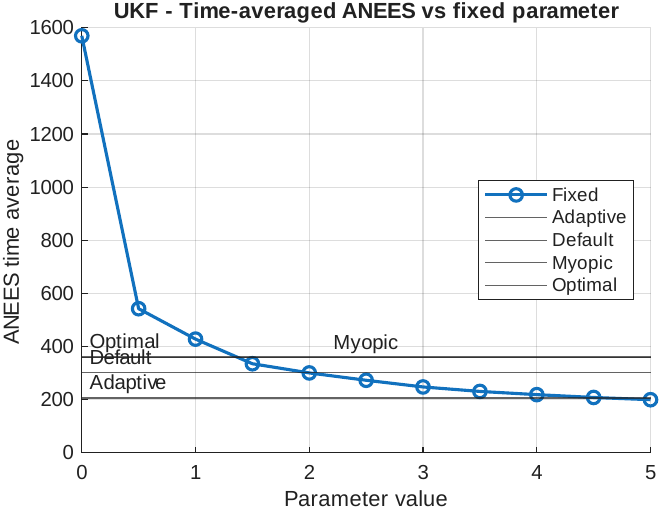}
    \caption{Time averaged \ac{anees} for \ac{ukf} and \ac{ungm}.}
    \label{fig:ungm_ukf_anees}
\end{figure}

\begin{figure}
    \centering
    \includegraphics[width=0.8\linewidth]{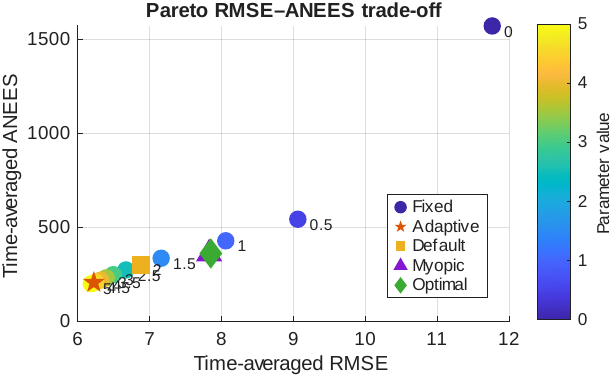}
    \caption{Time averaged \ac{anees} vs. \ac{rmse} for \ac{ukf} and \ac{ungm}.}
    \label{fig:ungm_ukf_rmse_anees}
\end{figure}

\begin{figure}
    \centering
    \includegraphics[width=0.7\linewidth]{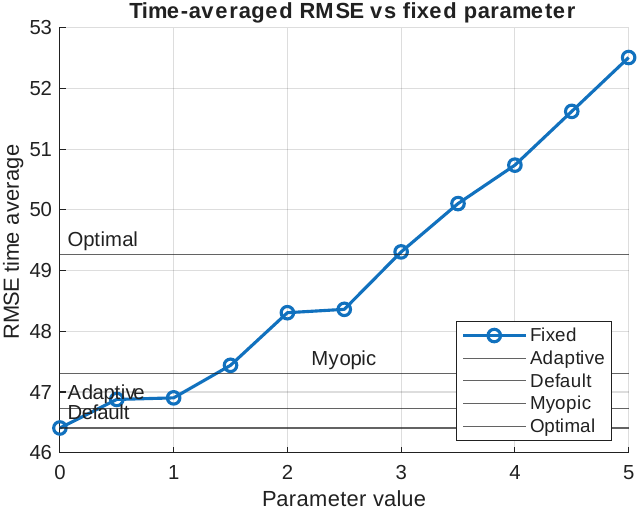}
    \caption{Time averaged \ac{rmse} for \ac{ukf} and \ac{ctm}.}
    \label{fig:ctm_ukf_rmse}
\end{figure}

\begin{figure}
    \centering
    \includegraphics[width=0.7\linewidth]{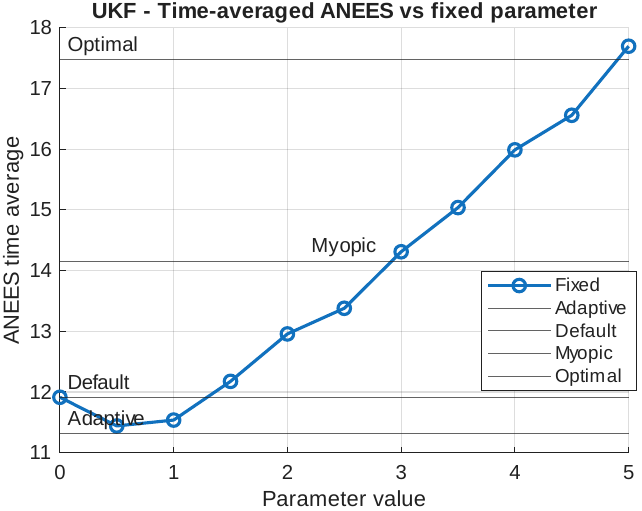}
    \caption{Time averaged \ac{anees} for \ac{ukf} and \ac{ctm}.}
    \label{fig:ctm_ukf_anees}
\end{figure}

\begin{figure}
    \centering
    \includegraphics[width=0.8\linewidth]{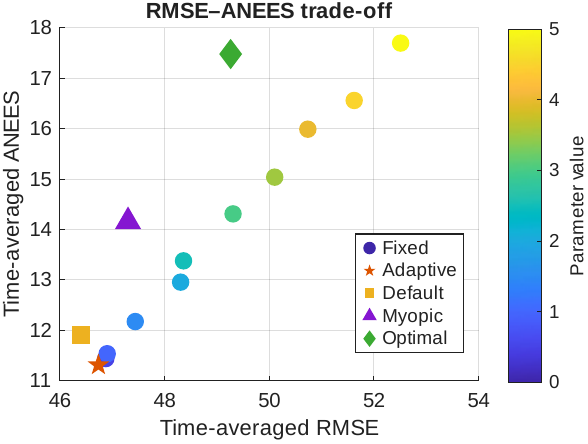}
    \caption{Time averaged \ac{rmse} vs. \ac{anees} for \ac{ukf} and \ac{ctm}.}
    \label{fig:ctm_ukf_rmse_anees}
\end{figure}

Time averages of \ac{rmse} and \ac{anees} for the \ac{ungm} and \ac{ukf} are given in Figures~\ref{fig:ungm_ukf_rmse} and~\ref{fig:ungm_ukf_anees}.  
In addition, the \ac{rmse}–\ac{anees} plot~\ref{fig:ungm_ukf_rmse_anees} visualizes the trade-off between estimation accuracy and statistical consistency by mapping each filter configuration to a point whose horizontal position reflects time-averaged \ac{rmse} and vertical position reflects time-averaged \ac{anees}.
Analogously, the results for the \ac{ctm} and \ac{ukf} are given in Figures~\ref{fig:ctm_ukf_rmse}, \ref{fig:ctm_ukf_anees}, and~\ref{fig:ctm_ukf_rmse_anees}.

In both \ac{ungm} and \ac{ctm} problems, the proposed parameter adaptation achieves the best consistency in terms of \ac{anees} compared to the myopic, optimal, and default baselines.
In terms of \ac{rmse}, the proposed parameter adaptation achieves the lowest values compared to the myopic, optimal, and default baselines.
The same estimate accuracy was achieved by the \ac{ukf} with one of the fixed parameter values.
This value, however, was not known in advance.
Also, it must be noted that the parameter adaptation used the \ac{nis} cost that values consistency in the measurement space, which is rather closer to state consistency assessed by \ac{anees} than to accuracy \ac{rmse}.
Both \ac{ungm} and \ac{ctm} problems also demonstrated better performance of parameter adaptation with longer horizon (despite a relatively low forgetting factor $\gamma=0.5$) compared to the myopic baseline that corresponds to $\gamma=0$.

The \ac{sif} was applied to \ac{ctm} only since \ac{ungm} has one-dimensional state and \ac{sif} reduces to an \ac{mc} \ac{kf} in such a case.
The results are given in Table~\ref{tab:sif_ cost_comparison}. 
In this case, only two baselines are used: the one with myopic optimization and the default parameter. 
The reason is that, in this case, the computational cost is included in the cost, and the computational cost weight can be adjusted to value either quality or cost.
A low weight leads to an optimum $N_\text{it}$ close to the maximum number of iterations, while a high weight results in the optimum  $N_\text{it}$ near the minimum number of iterations.
The results again indicate that parameter adaptation leads to better accuracy and consistency than the myopic and fixed-parameter baselines.
The last row of the table also contains the time-average value of the corresponding cost evaluated myopically (i.e., over a single time step).
In contrast to \ac{rmse} and \ac{anees}, this cost also accounts for computational costs, as shown in the column labels.
Here, the parameter adaptation also yields values smaller than those from myopic adaptation.
While this situation may seem counterintuitive, it is caused by the myopic setting selecting the best value of the parameter, taking into account a single time instant, without regard for the future.
Such greedy optimization cannot achieve the minimum cost over a long horizon, which is 500 steps in this case.




\begin{table*}[t]
\centering
\caption{\ac{sif} for \ac{ctm}: Time-averaged performance \ac{rmse}, \ac{anees}, and instantaneous cost $\bL$ for three cost functions.}
\label{tab:sif_ cost_comparison}
\setlength{\tabcolsep}{5pt}
\renewcommand{\arraystretch}{1.2}
\begin{tabular}{l|ccc|ccc|ccc}
\toprule
 & \multicolumn{3}{c|}{$\bL_\text{stateInnov}+\tfrac{1}{50}N_\text{it}$} 
 & \multicolumn{3}{c|}{$\bL_\text{logmaxNIS}+\tfrac{1}{50}N_\text{it}$}
 & \multicolumn{3}{c}{$\bL_\text{NIS}+\tfrac{1}{50}N_\text{it}$} \\
Metric 
 & Adaptive & Fixed & Myopic
 & Adaptive & Fixed & Myopic
 & Adaptive & Fixed & Myopic \\
\midrule
Time Avg. RMSE  & 5.630 & 6.013 & 5.778 & 6.866 & 6.462 & 6.771 & 6.359 & 6.537 & 6.646 \\
Time Avg. ANEES & 14.369 & 18.389 & 16.987 & 7.379 & 8.024 & 8.195 & 7.051 & 8.271 & 7.483 \\
Time Avg. cost  & 49.872 & 62.251 & 51.770 & 1.809 & 1.881 & 1.906 & $6.140\times10^7$ & $6.356\times10^7$ & $1.071\times^8$ \\
\bottomrule
\end{tabular}
\end{table*}

\section{Conclusion}\label{sec:conclusion}
In this paper, we investigated parameter adaptation for nonlinear state estimation from a decision-theoretic perspective, treating it as a sequential optimization over the filter parameter values. 
By integrating reinforcement learning with classical filtering, the proposed approach enables a non-myopic parameter adaptation that explicitly balances estimation accuracy, consistency, and computational costs. 
Empirical results from various cost forms show that adaptive parameter policies consistently outperform traditional fixed-parameter baselines and enhance myopic optimization.
These findings indicate that treating filter parameter adaptation as a sequential decision problem offers a systematic and practical approach to robustly deploying nonlinear estimators.
Out-of-distribution robustness and few-shot retuning of the learned policy are left for future work.


\end{document}